\documentclass[prr,notitlepage,superscriptaddress,amsmath,amssymb,twocolumn,longbibliography]{revtex4-2}
\usepackage{graphicx}
\usepackage{xcolor}
\usepackage{siunitx}

\begin{document}

 \title{Characterization of helical Luttinger liquids\\
 in microwave stepped-impedance edge resonators}

\author{Alexandre Gourmelon}
\author{Hiroshi Kamata}
\author{Jean-Marc Berroir}
\author{Gwendal F\`eve}
\author{Bernard Pla\c cais}
\affiliation{Laboratoire de Physique de l'\'Ecole Normale Sup\'erieure, ENS, PSL Research University, CNRS, Sorbonne Universit\'e, Universit\'e Paris Diderot, Sorbonne Paris Cit\'e, 24 rue Lhomond, 75231 Paris Cedex 05, France}
\author{Erwann Bocquillon}
\email{erwann.bocquillon@ens.fr}
\affiliation{Laboratoire de Physique de l'\'Ecole Normale Sup\'erieure, ENS, PSL Research University, CNRS, Sorbonne Universit\'e, Universit\'e Paris Diderot, Sorbonne Paris Cit\'e, 24 rue Lhomond, 75231 Paris Cedex 05, France}

\date{\today}

\begin{abstract}
Coulomb interaction has important consequences on the physics of quantum spin Hall edge states, weakening the topological protection via two-particle scattering and renormalizing both the velocity and charge of collective plasmon modes compared to that of free electrons. Despite these effects, interactions remain difficult to quantify. We propose here simple and robust edge resonator geometries to characterize Coulomb interaction by means of high-frequency measurements. They rely on a transmission line approach, and take advantage of the impedance mismatch between the edge states and their microwave environment.
\end{abstract}

\pacs{}
\keywords{}

\maketitle

{\bf
}

 The helical edge states of 2D topological insulators, which exhibit the quantum spin Hall (QSH) effect \cite{konig2007}, offer an exciting playground for exotic topological physics such as spin-polarized edge transport \cite{roth2009,knez2011,brune2012} or topological superconductivity \cite{bocquillon2016,deacon2017}. Another point of interest is the study of Coulomb interactions, which are particularly prominent in one-dimensional systems. In 2D topological insulators, one-particle scattering is in principle suppressed by time-reversal symmetry. However two-particle interactions are constrained but not forbidden \cite{wu2006}, and could have important consequences. They could for example generate back-scattering via diverse mechanisms \cite{maciejko2009,crepin2012,hsu2017}. In particular, inter-channel interaction leads to a modification of the charge but also the spin polarization of the edge plasmons, thus potentially degrading the performance of spin(orbi)tronics devices \cite{pesin2012, hofer2014} and obscuring the observation of topological transport.
 
At a microscopic level, Coulomb interaction strongly alters the dynamics of QSH helical edge states. First, the electron Fermi velocity $v_F$ is renormalized to a larger value $v$ under the action of Coulomb repulsion. Second, as charges propagate in one edge channel, inter-channel interaction drags a small amount of charge in the channel of opposite direction, leading to a non-trivial effective charge of the plasmon modes, reduced from that of the electron by a factor $\sqrt{K}\in[0,1]$, where $K$ is called the Luttinger parameter. The two parameters $K$ and $v$ fully characterize the dynamical properties of helical edge channels, but remain experimentally hardly accessible. In particular, $K$ is quite elusive: the dc conductance of an ohmically contacted 1D system does not depend on $K$ \cite{safi1995,maslov1995}, which is thus primarily accessed via power law exponents (tunneling density of states \cite{stuhler2019}, temperature dependence of the conductance \cite{li2015,sato2019}) or current correlations \cite{trauzettel2004,safi2008}. The ac conductance also inherits a $K$ dependence, but studies have concentrated mostly on the low-frequency regime \cite{blanter1998}, which do not capture the velocity $v$, on more complex setups \cite{muller2017} or on chiral edge modes \cite{kamata2014,brasseur2017}.

High-frequency experiments have proven very adequate in the context of the quantum Hall effect to investigate chiral edge magnetoplasmons \cite{ashoori1992,zhitenev1993,sukhodub2004,gabelli2007,kamata2010,hashisaka2013,kumada2011,bocquillon2013,kumada2014a,freulon2015,mahoney2017,mahoney2017a}. In the same spirit, we establish in this letter that the scattering of microwaves on capacitively coupled resonators offers a straightforward characterization of both $v$ and $K$. The high impedance edge channels are then confined between low impedance input and output circuitry. The impedance mismatch generates reflection at each interface, in a geometry analogous to so-called stepped-impedance resonators, heavily used in acoustics or microwave design. This geometry is advantageous: i) the use of high frequencies ($>\SI{10}{\giga\hertz}$) transport allows for using short devices ($\sim \SI{10}{\micro\meter}$) in the ballistic limit ii) the capacitive coupling circumvents complications due to ohmic contacts: contact resistances which bring dissipation \cite{viola2014}, as well as a complex behavior in the GHz range \cite{wilmart2018} iii) the geometry can be analyzed in terms of microwave networks \cite{pozar2012}, combining simple experimental setups \cite{pallecchi2011,inhofer2017,dartiailh2020}, and straightforward interpretation.


\paragraph{Edge channels and equivalent transmission lines --}

\begin{figure}[ht]
\centerline{\includegraphics[width=8.7cm]{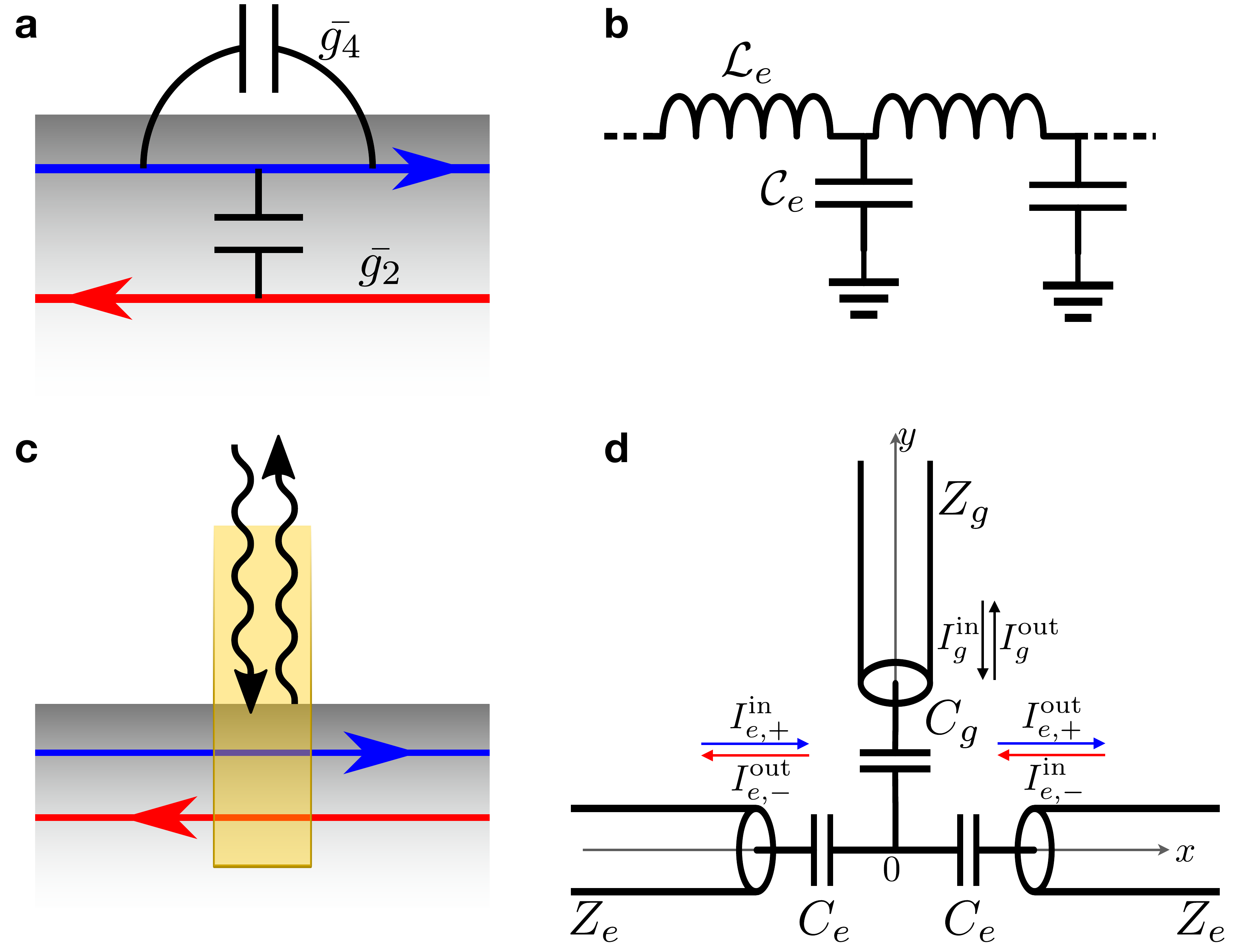}}
\caption{{\bf Schematics of the setup:} a) Schematic view of inter- and intra-channel Coulomb interaction in helical edge states (in red and blue solid lines), with their respective coupling constants $\bar{g_2}$ and $\bar{g_4}$. b) Equivalent transmission line representation, with line inductance $\mathcal{L}_e$ and line capacitance $\mathcal{C}_e$. c) Schematic view of a gate coupling capacitively to helical edge states d) Equivalent transmission line representation, with geometric capacitance $C_g$ and quantum capacitance $C_e$.
}\label{figure:Lines}
\end{figure}

In this section we introduce the bosonized Luttinger representation of the helical edge states, and establish its relation to a microwave transmission, following previous works \cite{tarkiainen2001,burke2002}. For helical edge channels, the bosonized Hamiltonian is formally identical to that of a spinless Luttinger liquid and reads \cite{wu2006,dolcetto2015}:
\begin{equation}
    H = \frac{v}{2\pi}\int dx [\frac{1}{K}(\partial_x\phi)^2 + K(\partial_x\theta)^2] - e U \partial_x\phi 
    \label{H_hLL}
\end{equation}
where the fields $\phi$ and $\theta$ fields are related to the right ($\phi_R$) and left ($\phi_L$) bosonic fields representing right and left-movers, with $\phi =  \phi_R + \phi_L,\  \theta = \phi_R - \phi_L$. $U$ is an external potential, in a minimal coupling approach. Coulomb interaction renormalizes the sound velocity of the collective excitations velocity $v$, and defines the Luttinger parameter $K$, which are expressed as a function of the Fermi velocity $v_F$, the inter-channel (resp. intra-channel) coupling constants $\bar{g_2}$ (resp. $\bar{g_4}$):
\begin{eqnarray}
   v &=& v_F\sqrt{(1+\bar{g_4}+\bar{g_2})(1+\bar{g_4}-\bar{g_2})}
   \label{v}\\
    K &=& \frac{\sqrt{1+\bar{g_4}-\bar{g_2}}}{\sqrt{1+\bar{g_4}+\bar{g_2}}}
    \label{K}
\end{eqnarray}
In the absence of external potential ($U=0$), this Hamiltonian is completely equivalent to that of a $LC$ distributed transmission line \cite{pozar2012} with line capacitance $\mathcal{C}_e$ and line inductance $\mathcal{L}_e$, such as the one sketched in Fig.\ref{figure:Lines}b:
 \begin{eqnarray}
      \mathcal{H} &=&  \frac{\rho_e^2}{2\mathcal{C}_e} + \frac{1}{2}\mathcal{L}_eI_e^2\\
       {\rm with\ }   \mathcal{C}_e &=& \frac{2K}{R_Kv}, \quad \mathcal{L}_e= \frac{R_K}{2Kv}
 \end{eqnarray}
where $\rho_e=\frac{e}{\pi}\partial_x \phi$ is the charge density and $I_e=\frac{evK}{\pi}\partial_x \theta$ the current flowing on the edge, and $R_K$ the quantum of resistance. This identification establishes the equivalence between a helical Luttinger liquid and a transmission line with characteristic impedance $Z_e$, and velocity $v$ given by \footnote{We note that this approach also includes gate screening effects, simply by renormalizing $\mathcal{C}_e$}:
 \begin{equation}
      Z_e = \sqrt{\frac{\mathcal{L}_e}{\mathcal{C}_e}} = \frac{R_K}{2K}, \qquad v=\frac{1}{\sqrt{\mathcal{L}_e\mathcal{C}_e}}
 \end{equation}
 This transmission line hosts two modes with linear dispersion and velocity $v$, propagating in opposite directions (denoted $\pm$). Here, one can interpret the factor $K$ as a reduced effective charge $\sqrt{K}e$ affecting $ \mathcal{L}_e, \mathcal{C}_e$ and in turn the line impedance $Z_e$. 



One can introduce the currents $I_{e}(x,t)$ and voltages $V_e(x,t)=\rho_e(x,t)/\mathcal{C}_e$ \cite{pozar2012,burke2002}:
\begin{eqnarray}
    V_{e}(x,t) &=& \int \frac{d\omega}{2\pi}\left( V_{e}^{-}e^{j\omega (t + \frac{\omega}{v} x)} + V_{e}^{+}e^{j\omega( t - \frac{\omega}{v} x)}\right)\\
    I_{e}(x,t) &=& \int \frac{d\omega}{2\pi}\left(-I_{e}^{-}e^{j\omega (t + \frac{\omega}{v} x)} + I_{e}^{+}e^{j\omega( t - \frac{\omega}{v} x)}\right)
\end{eqnarray}
such that  $Z_{e} = \frac{V_{e}^{\pm}}{I_{e}^{\pm}}$.

In this framework, the currents  $I_{e}^{\pm}$ carried by both modes $\pm$ then simply acquire a phase factor $s_\pm=e^{\pm i\frac{\omega}{v}(x-x')}$ as they propagate from point $x$ to $x'$.

\paragraph{Capacitive coupling in a transmission line approach --}

This transmission line approach allows for a simple description of elaborate geometries, in close analogy to microwave network analysis. Here we start by describing capacitive contacts. The latter are essentially dissipationless, thus circumventing the dissipation inherent to the Landauer contact resistance. Besides, such reactive contacts achieve a simple capacitive coupling even at high frequencies, unlike ohmic contacts which exhibit mixed capacitive and resistive behaviors \cite{wilmart2018}. They have been heavily used in quantum Hall high frequency experiments \cite{ashoori1992, kamata2014, feve2007, bocquillon2013} and have triggered recent interest in Hall gyrators and circulators \cite{viola2014,bosco2017,mahoney2017,mahoney2017a}
The capacitive contact is described as a three-port device coupling at $(x,y)=(0,0)$ the edge states (seen as two semi-infinite transmission lines of impedance $Z_e$, for $x>0$ and $x<0$) to the gate, seen as a semi-infinite transmission line of impedance $Z_g=\SI{50}{\ohm}$ along the $y>0$ axis), in the geometry shown in Fig.\ref{figure:Lines}d. The geometrical gate capacitance is noted $C_g$ while quantum capacitance effects are accounted for by $C_e$.
This point-like description is inspired by transmission line models but similar results have been obtained with a more elaborate plasmon distributed model, adding some finite size effects which are disregarded here (see more in section {\it Experimental considerations}).

Introducing the Fourier modes $I_{g}^\pm$ and $V_{g}^\pm$ for the gate transmission line, one obtains from Kirchhoff's laws at the coupling point $(x,y)=(0,0)$ the relations between input and output currents in each of the arms:
\begin{eqnarray}
     I_{g}^{\rm in} - I_{g}^{\rm out} &=& \sum_{i=\pm} \left( I_{e,i}^{\rm out} - I_{e,i}^{\rm in} \right)\\
  \forall\ i=\pm,\   \Lambda_e I_{e,i}^{\rm out} - \Lambda_e^* I_{e,i}^{\rm in } &=& \frac{C_{e}}{C_g} \left( \Lambda_g I_g^{\rm out}  - \Lambda_g^* I_g^{\rm in} \right)
\end{eqnarray}
where $I_g^{\rm in/out}=I_g^\pm |_{y=0^+}$,  $I_{e,\pm}^{\rm in}=I_e^\pm |_{x=0^\mp}, I_{e,\pm}^{\rm out}=I_e^\pm |_{x=0^\pm}$ (see Fig.\ref{figure:Lines}d), and $\Lambda_e(\omega) = 1 + j\omega Z_{e} C_{e}$, $\Lambda_g(\omega) = 1 + j\omega Z_{g} C_{g}$. Solving these three equations yield the scattering matrix elements, relating outgoing to ingoing {\sl currents} \footnote{To avoid confusions, we recall that, unlike here, scattering matrix are usually defined for modes or voltages \cite{pozar2012}} :
\begin{eqnarray}
    S_{gg} &=& \frac{1 + \frac{2C_{e}}{C_g}\frac{\Lambda_g^*(\omega)}{\Lambda_e(\omega)}}{1 + \frac{2C_{e}}{C_g}\frac{\Lambda_g(\omega)}{\Lambda_e(\omega)}}\\
    S_{g\pm} &=& \frac{2j\omega Z_eC_{e}\frac{1}{\Lambda_e(\omega)}}{1 + \frac{2C_{e}}{C_g}\frac{\Lambda_g(\omega)}{\Lambda_e(\omega)}}= \frac{Z_e}{Z_g}S_{\pm g}\\
     S_{\pm \pm} &=& \frac{\Lambda_e^*(\omega)}{\Lambda_e(\omega)}\frac{1+\frac{2\Lambda_g(\omega)}{\Lambda_e^*(\omega)}\frac{C_e}{C_g}}{1 + \frac{2\Lambda_g(\omega)}{\Lambda_q(\omega)} \frac{C_{e}}{C_X}}\\
    S_{\pm \mp} &=& \frac{C_{e}}{C_g}\frac{\Lambda_g(\omega)}{\Lambda_e^2(\omega)}\frac{2j\omega Z_e C_e}{1 + \frac{2C_{e}}{C_g}\frac{\Lambda_g(\omega)}{\Lambda_e(\omega)}}
\end{eqnarray}
\begin{figure}[ht]
\centerline{\includegraphics[width=8cm]{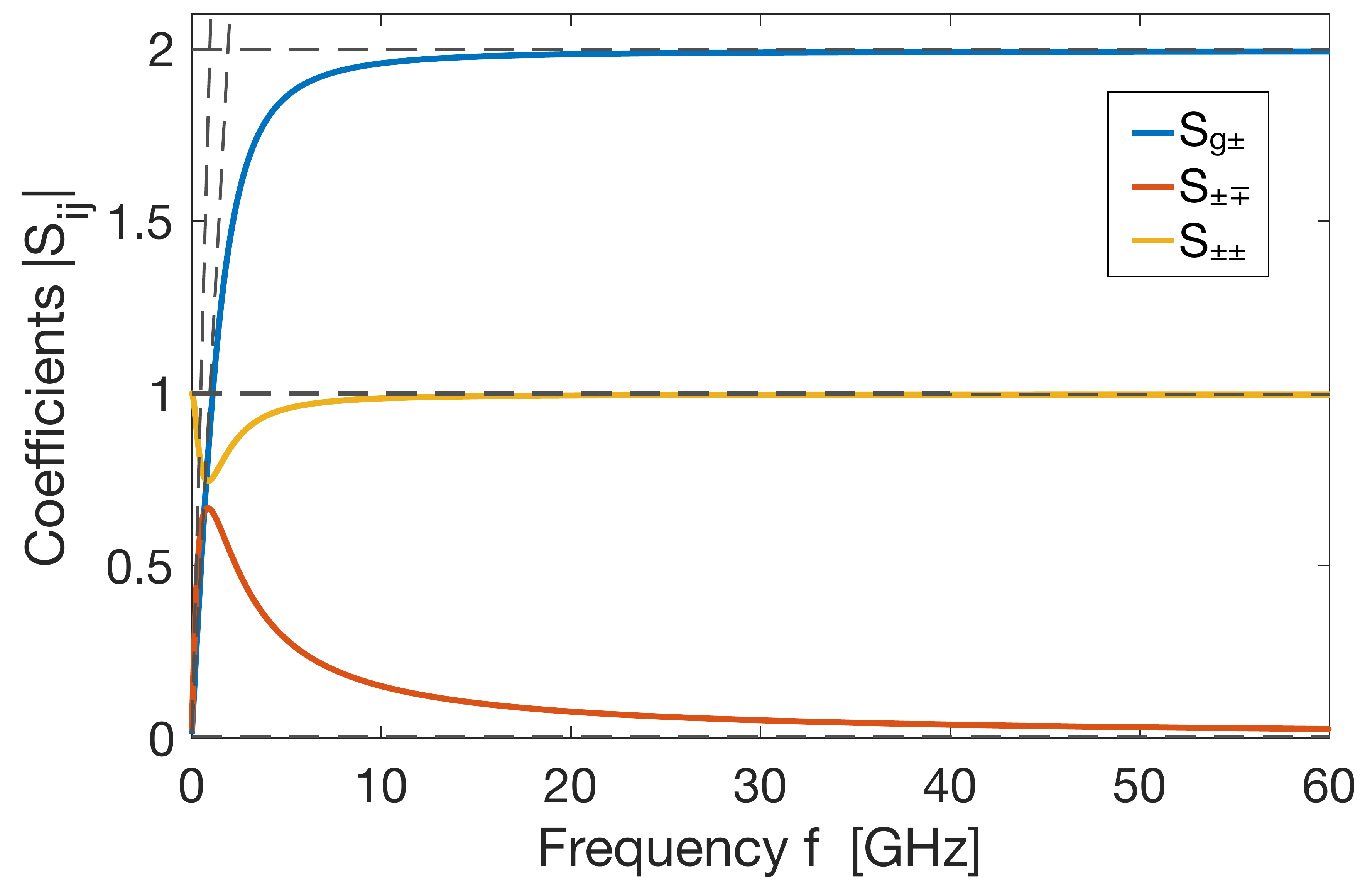}}
\caption{{\bf Scattering coefficients of a capacitive contact:} $|S_{g\pm}|$ is represented a a blue line, $|S_{\mp\pm}|$ as a red line, and $|S_{\pm\pm}|$ as a yellow line. For each coefficient, the grey dashed lines indicate the low and high frequency asymptotes, obtained analytically. The graphs are obtained with parameters $C_e=\SI{16}{\femto\farad}, C_g=\SI{8}{\femto\farad}, K=0.5$.}
\label{figure:TLM_scattering}
\end{figure}

The modulus of $S_{g\pm}, S_{\pm\pm}$ and $S_{\pm\mp}$ are plotted in Fig.\ref{figure:TLM_scattering} as function of frequency $f=\frac{\omega}{2\pi}$, for a realistic set of parameters (see {\it Experimental considerations}). In these coefficients, two RC time-scales appear, namely the gate one $Z_g C_g$ and that of the edge states $Z_e C_e$, defining two regimes: low frequencies for which $\omega\ll 1/Z_g C_g, 1/Z_e C_e$, and high frequencies with $\omega\gg  1/Z_g C_g, 1/Z_e C_e$.
In the low frequency regime, the capacitive elements $C_g,C_e$ play the dominant role. We in particular observe that $S_{g\pm}\simeq 2j\omega Z_e C_e$, $S_{\pm\mp}\simeq j\omega \frac{C_e}{C_g}C_t$, with $C_t=\frac{2C_eC_g}{C_g+2C_e}$ the total capacitance. 

In contrast, at high frequency, the capacitors are transparent, and do not play any role. The scattering elements are thus dominated by the impedance mismatch \cite{pozar2012} between the edge states (impedance $Z_e$) and the gate (impedance $Z_g$) with for instance $S_{g\pm}=\frac{2Z_e}{Z_e+2Z_g}\simeq 2$. This high frequency limit of transport and the role of the low impedance environment are both often disregarded but play here a crucial role. Indeed, as we will see below, their interplay through impedance mismatch provides a very straightforward approach to measuring $Z_e$ and thus $K$.

\paragraph{Linear and ring resonators --}

\begin{figure}[ht]
\centerline{\includegraphics[width=9cm]{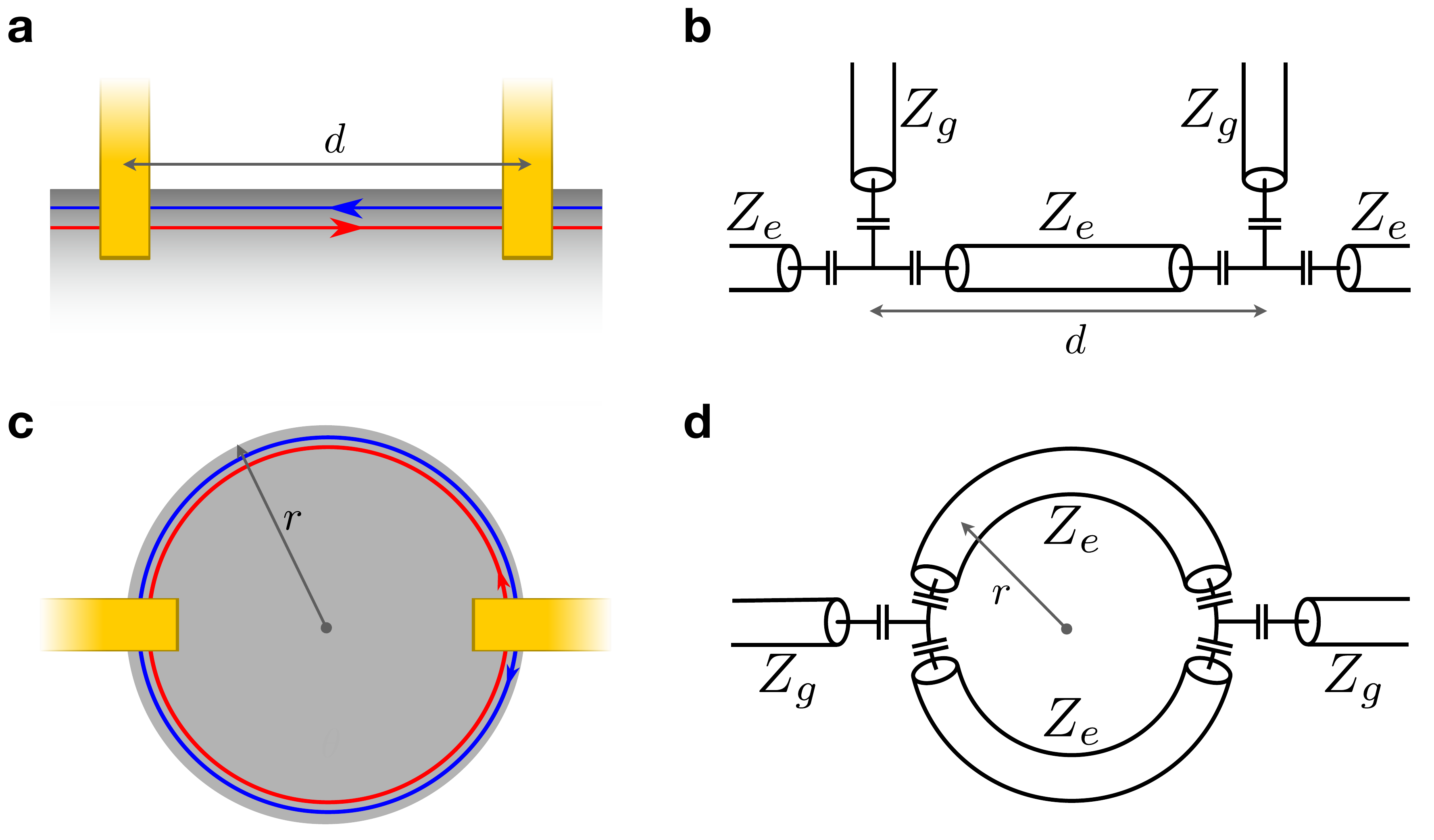}}
\caption{{\bf Geometry of the linear and ring resonators:} a, c) Schematic view of a linear (a) and ring (c) resonator for the edge states of the QSH effect (in red and blue solid lines). Capacitive contacts are depicted in yellow. b,d) Equivalent transmission line representation of the linear (b) and ring (d) resonators.}
\label{figure:ResoGeom}
\end{figure}

In this section, we introduce the geometry of linear and ring "resonators" in which the two edge states are confined between two capacitive contacts (Fig.\ref{figure:ResoGeom}). In the linear resonator (Fig.\ref{figure:ResoGeom}a), the contacts are separated by a distance $d$ and we assume that the regions outside the contacts are semi-infinite, and sink all incoming wave. In the ring geometry (Fig.\ref{figure:ResoGeom}c), the edge states circulate along a disk-shaped mesa of radius $r$, with the two contacts placed symmetrically on either side of the disk.
Each structure forms a two-port device, with equivalent transmission line representations depicted in Fig.\ref{figure:ResoGeom}b and \ref{figure:ResoGeom}d. The schemes in Fig.\ref{figure:ResoGeom}b and \ref{figure:ResoGeom}d highlight the analogy with stepped impedance resonators, with regions of impedance $Z_e$ comprised between ports of lower impedance $Z_g$. Using the scattering matrix of the free QSH edge and of the capacitive contact, we compute the transmission $T$ of each device and analyze it below. The reflexion coefficient on each contact is disregarded as it is mostly dominated by $S_{gg}\simeq 1$, and we assume for now that the devices are ballistic.

Starting with the linear resonator, summing the contributions of all waves, one obtains the transmission:
\begin{eqnarray}
T &=& \frac{S_{g+}S_{+g}e^{-j\omega\frac{d}{v}}}{1 - S_{+-}S_{-+}e^{-2j\omega\frac{d}{v}}}= \frac{Z_g}{Z_e} \frac{S^2_{g+}e^{-j\omega\frac{d}{v}}}{1 - S_{+-}^2e^{-2j\omega\frac{d}{v}}}
\end{eqnarray}

\begin{figure*}[ht]
\centerline{\includegraphics[width=17cm]{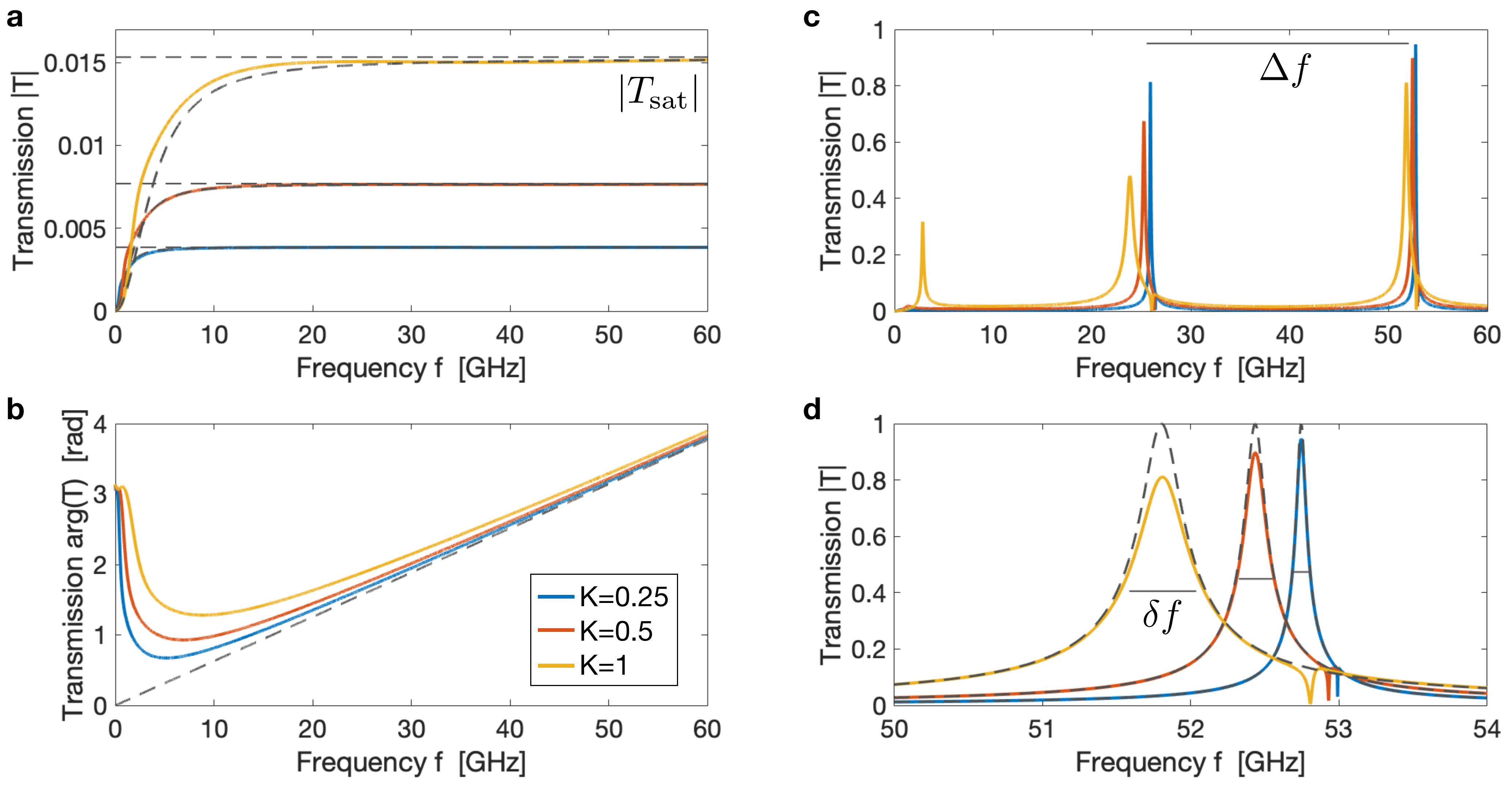}}
\caption{{\bf Transmission for the linear and ring resonators:} a,b) For $K=0.25, 0.5, 1$, the modulus $|T|$ (panel a) and phase $\arg(T)$ (panel b) of the transmission of the linear resonators are plotted as solid colored lines. In panel a), the dashed lines represent the approximation $\frac{Z_g}{Z_e} |S_{g+}|^2$ and the high frequency asymptote $|T_{\rm sat}|$, allowing to determine $K$. In panel b), the dashed line represents the asymptote $\omega d/v$, giving access to $v$.  c,d) For $K=0.25, 0.5, 1$, the modulus $|T|$ of the transmission of the ring resonators are plotted as solid colored lines. In panel c), the solid grey line represents the interval $\Delta f$ between two resonances. In panel d), the dashed lines represents the approximation given in Eq.(\ref{eq:approx_ring}), while the solid grey lines represent the peak widths $\delta f$, allowing to determine $K$. All graphs are obtained with parameters $C_e=\SI{16}{\femto\farad}, C_g=\SI{8}{\femto\farad}, v=\SI{1e6}{\meter\per\second}, d=\SI{10}{\micro\meter}, r=\SI{6}{\micro\meter}$.}
\label{figure:ResonatorsT}
\end{figure*}

$T$ exhibits the familiar form of a Fabry-P\'erot resonator: the numerator reflects the direct path from one gate to the other, while the denominator encodes the multiple round-trips in the cavity. The modulus $|T|$ and phase $\arg(T)$ are plotted in Fig. \ref{figure:ResonatorsT} (panels a and b) for $K=0.25, 0.5$ and 1. We first note, that, since $S_{+-}$ remains small, the FP oscillations are quite weak, and most of the excitation signal leaks to the regions located beyond both capacitive contacts. The transmission $T$ is thus governed by its numerator (plotted as a grey dashed line in Fig.\ref{figure:ResonatorsT}a). At low frequencies (for $\omega\ll  1/Z_g C_g, 1/Z_e C_e$), $T$ is quite low, dominated by the succession of two capacitive elements $S_{g+}^2$, with $|T|\propto \omega^2$. The physics of interactions remains mostly inaccessible.  On the opposite, at high frequencies, $|T|$ saturates at a maximum value $|T_{\rm sat}|$ imposed by the saturation of $S_{g+}$, with $|T_{\rm sat}|=\frac{4Z_g}{Z_e}=\frac{8KZ_g}{R_K}$. In this regime, $S_{g+}$ is real, such that $\arg(T)\simeq\frac{\omega d}{v}$.
This simple geometry thus allows for a very direct readout of the velocity $v$ (from $\arg(T)$) and of the Luttinger interaction parameter $K$ (via $|T_{\rm sat}|$).

We now move on to the case of the ring resonator. In this case, many waves, scattering on either of the contacts, interfere and contribute to the output current. The transmission $T$ can be analytically derived in full generality and reads:
\begin{widetext}
\begin{eqnarray}
T &=& \frac{2Z_g}{Z_e} \frac{S_{g+}^2e^{-j\omega\frac{\pi r}{v}}\Big(1-e^{-2j\omega\frac{d}{v}}(S_{++}^2-S_{+-}^2)\Big)}{\Big(1 - 2e^{-j\omega\frac{d}{v}}S_{+-}-e^{-j\omega\frac{2\pi r}{v}}(S_{++}^2-S_{+-}^2)\Big)\Big(1 + 2e^{-j\omega\frac{\pi r}{v}}S_{+-}-e^{-j\omega\frac{2\pi r}{v}}(S_{++}^2-S_{+-}^2)\Big)}
\end{eqnarray}
\end{widetext}
As can be seen from Fig.\ref{figure:ResonatorsT}c and \ref{figure:ResonatorsT}d, $T$ exhibits many features, with a quite complex behavior. Nonetheless, in the high frequency regime $\omega\gg  Z_g C_g, Z_e C_e$, the different $S_{ij}$ coefficients are all real, and $S_{+-}$ is rather small, so that $T$ can be approximated by the following formula (plotted as a dashed grey line):
\begin{eqnarray}
T &=& \frac{2Z_g}{Z_e} \frac{S_{g+}^2e^{-j\omega\frac{\pi r}{v}}}{1 -e^{-j\omega\frac{2\pi r}{v}}S_{++}(S_{++}+2S_{+-})}
\label{eq:approx_ring}
\end{eqnarray}
In this regime, one again recognizes a familiar Fabry-P\'erot resonator, with an effective "mirror" reflection coefficient $\Gamma=S_{++}(S_{++}+2S_{+-})$. Equidistant resonance peaks are separated by $\Delta f=\frac{v}{2\pi r}$ the round-trip frequency, allowing for a simple extraction of $v$. Here, losses are ignored, and the geometry is closed. As a consequence, the amplitude of the peaks tends to 1 in the high frequency regime, regardless of parameters. However, the impedance $Z_e$ acts on the value of $S_{\pm\pm}$, hence on the reflexion $\Gamma$ which determines the peak width: we find that the width of the peaks (FWHM) is given by $\delta f=\frac{v}{2\pi r}\frac{16\sqrt{3} K Z_g}{\pi R_K}$, allowing for determining $K$. One can also simultaneously evaluate the relatively high quality factor $Q=\frac{\Delta f}{\delta f}=\frac{\pi R_K}{16\sqrt{3} K Z_g}\simeq \frac{60}{K}$.

\begin{figure}[ht]
\centerline{\includegraphics[width=8cm]{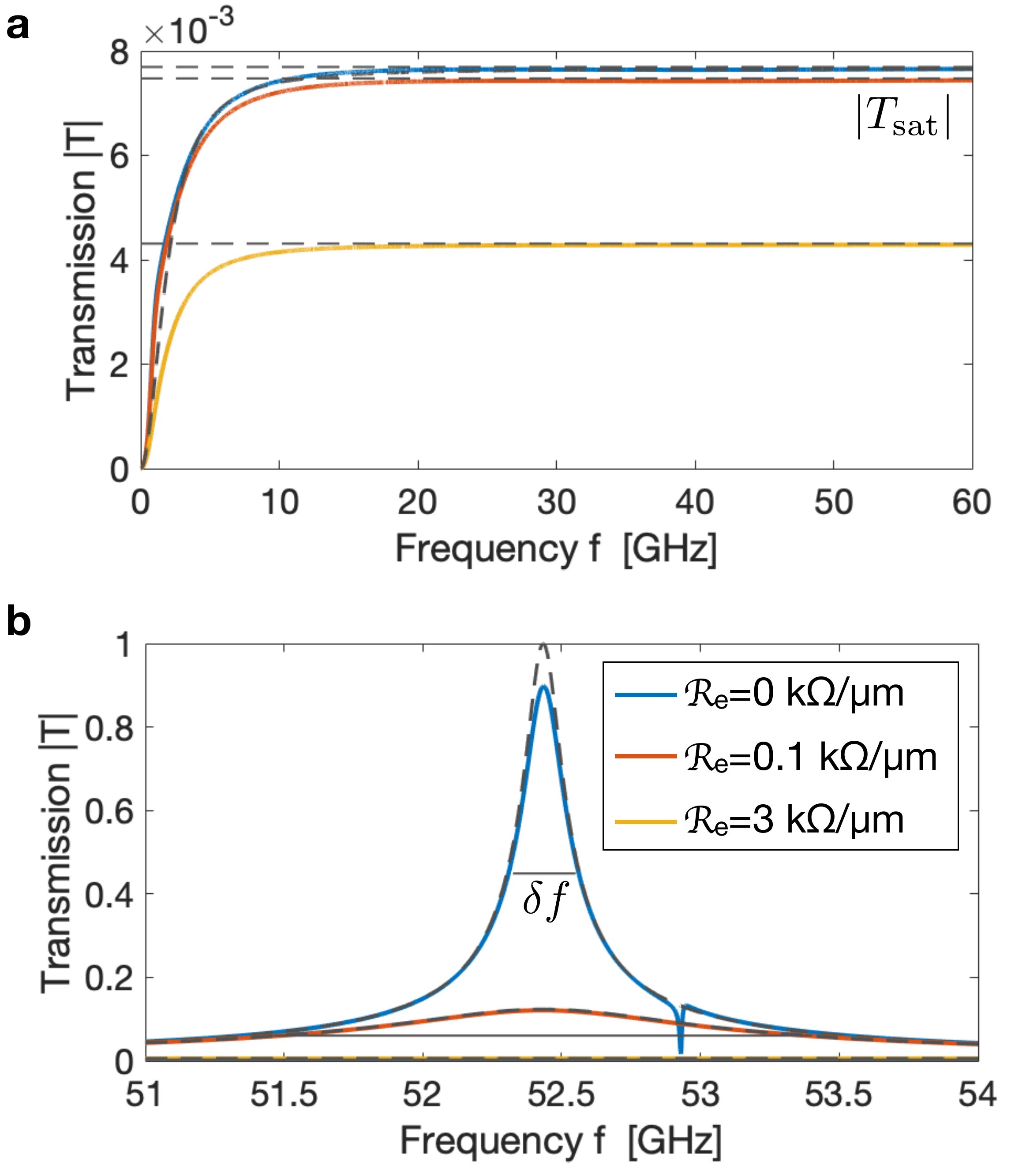}}
\caption{{\bf Effect of losses on the transmission:} a) For $\mathcal{R}_e=0, 0.1$ and $\SI{3}{\kilo\ohm\per\micro\meter}$, the modulus $|T|$ of the transmission of the linear resonator are plotted as solid colored lines. The dashed lines represent the high frequency asymptote $|T_{\rm sat}|$. b) For $\mathcal{R}_e=0, 0.1, \SI{3}{\kilo\ohm\per\micro\meter}$, the modulus $|T|$ of the transmission of the ring resonators are plotted as solid colored lines. The dashed lines represents the approximation given in Eq.(\ref{eq:approx_ring}), while the solid grey lines represent the peak widths $\delta f$. All graphs are obtained with parameters $C_e=\SI{16}{\femto\farad}, C_g=\SI{8}{\femto\farad}, v=\SI{1e6}{\meter\per\second}, K=0.5, d=\SI{10}{\micro\meter}, r=\SI{6}{\micro\meter}$.}
\label{figure:Losses}
\end{figure}

\paragraph{Beyond the ballistic limit --}
The devices previously introduced provide direct simultaneous access to $v$ and $K$. One can however wonder how scattering in the channels could alter the previous results. In this framework of transmission lines, a basic model of scattering is proposed via an additional line resistance $\mathcal{R}_e$, in series with $\mathcal{C}_e$. To lowest order \cite{pozar2012}, the propagation term $s_\pm$ then acquires an additional exponential term $e^{-\frac{\mathcal{R}_e}{2Z_e}d}$. It is then straightforward to show that the linear resonator exhibits a modified saturation value $|T_{\rm sat}|=\frac{8KZ_g}{R_K}e^{-\frac{\mathcal{R}_e d}{2Z_e}}$, while the phase $\arg(T)$ is not modified. The change of $|T_{\rm sat}|$ is illustrated in Fig.\ref{figure:Losses}a, for loss parameters recently measured in HgTe layers \cite{lunczer2019} for which $\mathcal{R}_e$ ranges from $\sim 0.1$ to $\sim\SI{3}{\kilo\ohm\per\micro\meter}$.
In the ring resonator, the FP peaks are more strongly modified. They have a decreased maximum transmission $|T_{\rm max}|<1$ and an increased width $\delta f$ reading:
\begin{eqnarray}
|T_{\rm max}|&=&\frac{8Z_g e^{-\frac{\pi\mathcal{R}_e r}{Z_e}}}{Z_e(1-e^{-\frac{\pi\mathcal{R}_e r}{Z_e}})+Z_g(1+e^{-\frac{\pi\mathcal{R}_e r}{Z_e}})}\\
\delta f&=&\frac{v\sqrt{3}}{2\pi^2 r}\big[(1+\frac{16K Z_g}{R_K})e^{\frac{\pi\mathcal{R}_e r}{Z_e}}-1\big]
\end{eqnarray}
As a consequence, in Fig.\ref{figure:Losses}b, we observe that a finite $\mathcal{R}_e$ strongly suppresses the Fabry-P\'erot resonances in the ring resonator. From Eq.\ref{eq:approx_ring}, one notices that for large $\mathcal{R}_e$, the featureless transmission $T$ is then analogous to that of the linear resonator, with $|T|\to\frac{16KZ_g}{R_K}e^{-\frac{2\pi r\mathcal{R}_e}{2Z_e}}$.
In any case, measurements of samples with different travel lengths should be employed to carefully take this effect into account.

\paragraph{Experimental considerations --} Finally, we review the conditions required to perform the experiments. 
First, we note the impedance mismatch which allows for determining $K$ yields rather small values of $|T_{\rm sat}|\simeq 0.005\simeq \SI{-45}{\decibel}$. Though small, these values are customary when measuring edge states in the microwave regime \cite{dartiailh2020} and remain experimentally measurable with standard Vector Network Analyzers (VNAs).

Second, we stress that we have restricted this study to short distances (here $d=\SI{10}{\micro\meter}, r=\SI{6}{\micro\meter}$) so that the edge states remain as close as possible to the ballistic limit, usually 5 to \SI{10}{\micro\meter} \cite{du2015, bendias2018, lunczer2019}. Besides, in a recent study of HgTe quantum wells \cite{dartiailh2020}, we have measured the quantum capacitance, which remains large even in the topological gap. From the data, we estimate that $v\gtrsim v_F \simeq\SI{1e6}{\meter\per\second}$, $C_g\simeq\SI{8}{\femto\farad}$ and $C_e=\SI{16}{\femto\farad}$ for a gate with a area $\mathcal{A}=\SI{2}{\square\micro\meter}$. Given these parameters, the high frequency range of the study identified in Figs. \ref{figure:TLM_scattering} and \ref{figure:ResonatorsT} lies beyond \SI{20}{\giga\hertz}. This is challenging but not inaccessible thanks to the development of cryogenic microwave probe stations. They allow measurements down to \SI{4}{\kelvin} and up to \SI{67}{\giga\hertz} \cite{graef2018,graef2019a}, with accurate in-situ microwave calibration. Besides, these parameters ensure that the point-like description of the gate coupling is sufficient, as the finite length of the gate $\sqrt{\mathcal{A}}$ is always much smaller than the plasmon wavelength in the whole frequency range, i.e. such that $\sqrt{\mathcal{A}}\ll v/f$.

In other systems, with smaller velocities \cite{du2017}, or when quantum and geometric capacitances are smaller, larger gates may be necessary. In that case, the transmission line approach with point-like contacts between gate and edge states in principle fails. A long-range distributed gate coupling can be worked out based on Ref. \onlinecite{degiovanni2009}. The results described here can then be adapted to take into account the additional effects (propagation, finite-size effects) arising \cite{ota2018}, and the validity of our approach is not questioned.

Finally, we point out that the presence of a metallic top-gate (which screens interactions) and its distance to the helical edge channels may significantly modify the value of $v$ and $K$. Several configurations (back gate, top gate, resistive top gate \cite{kumada2020}) could thus be tested. In combination with DC measurements, one could also assess the influence of Coulomb interactions on scattering in helical edge channels, as suggested by many theoretical works \cite{maciejko2009,crepin2012,hsu2017}.

\section*{Conclusion}
As a conclusion, we have used the analogy between helical Luttinger liquids of a QSH insulator and microwave transmission lines to develop simple models of helical Luttinger and their coupling to local capacitive contacts. We have combined these building blocks in two different resonator geometries, and have shown that the measurements of the microwave transmission coefficient $T$ allows for a very natural determination of the velocity $v$ and Luttinger parameter $K$. Thus, a full characterization of Coulomb interaction can be obtained, shining new light on its consequences on the dynamics and back-scattering in helical edge states.
The challenge resides in the use of high frequencies ($>\SI{10}{\giga\hertz}$), now readily accessible in cryogenic microwave probe stations, in combination with conventional VNAs.\\ \indent
We believe that the general framework developed here for helical liquids can be extended to other types of interaction 1D systems, such as the chiral edge states of the integer and fractional quantum Hall effects (though the geometries studied in this article are irrelevant). Future developments will thus aim at proposing geometries unveiling the exact nature of modes (charged vs neutral modes, Majorana edge states, etc.) and better understand the effects of edge reconstructions.

\begin{acknowledgements}

\section*{Data availability}
The data sets generated and analyzed during the current study are available from the corresponding author on reasonable request.

\section*{Acknowledgments}
This work has been supported by the ERC, under contract ERC-2017-StG 758077 "CASTLES". E.B. gratefully acknowledges enlightening discussions with R. Abdellatif.

\section*{Author contributions}
A.G., H.K. and E.B. proposed the model, and performed the analytical and numerical analysis. E.B. supervised the project. All authors participated to the analysis of the results and to the writing of the manuscript.

\end{acknowledgements}

\bibliography{Resonators_bib.bib}
\end{document}